%% file: main.tex
\documentclass[10pt]{article}

\usepackage{amsmath}
\usepackage{amssymb}

\usepackage{graphicx}

\usepackage{cite}
\usepackage{url}
\usepackage{color} 
\usepackage{censor}

\usepackage[labelfont=bf,labelsep=period,justification=raggedright]{caption}

\usepackage{setspace}
\usepackage{lineno}

\makeatletter
\renewcommand{\@biblabel}[1]{\quad#1.}
\makeatother

\date{}

\begin{document}

\begin{flushleft}
{\Large
\textbf{Privacy for Personal Neuroinformatics}
}
\\
Arkadiusz Stopczynski$^{1,2}$,
Dazza Greenwood$^{2}$,
Lars Kai Hansen$^{1}$,
Alex Sandy Pentland$^{2}$
\\
\bf{1} Technical University of Denmark\\
\bf{2} MIT Media Lab
\\
arks@dtu.dk, dazza@civics.com, lkai@dtu.dk, sandy@media.mit.edu
\end{flushleft}

\input{sections/abstract}


\input{sections/introduction}

\input{sections/methods}

\input{sections/legal}
\input{sections/discussion}


\bibliography{bibliography}



\end{document}

%% file: sections/abstract.tex
\section*{Abstract}

Human brain activity collected in the form of Electroencephalography (EEG), even with low number of sensors, is an extremely rich signal.
Traces collected from multiple channels and with high sampling rates capture many important aspects of participants' brain activity and can be used as a unique personal identifier, similarly to fingerprints, DNA, or a portrait.
The motivation for sharing EEG signals is significant, as a mean to understand the relation between brain activity and well-being, or for communication with medical services.
However, only a small part of the brain activity is under voluntary control, thus the information revealed by EEG may largely be unknown to the user.
As the equipment for such data collection becomes more available and widely used, the opportunities for using the data are growing; at the same time however inherent privacy risks are mounting.
The same raw EEG signal can be used for example to diagnose mental diseases, find traces of epilepsy, and decode personality traits.
The current practice of the informed consent of the participants for the use of the data either prevents reuse of the raw signal or does not truly respect participants' right to privacy by reusing the same raw data for purposes much different than originally consented to.
This becomes even a bigger problem as the data lives on and new processing methods can extract information that was not deemed possible previously.

Here we propose an integration of a personal neuroinformatics system, Smartphone Brain Scanner, with a general privacy framework openPDS.
We show how raw high-dimensionality data can be collected on a mobile device, uploaded to a server, and subsequently operated on and accessed by applications or researchers, without disclosing the raw signal.
Those extracted features of the raw signal, called answers, are of significantly lower-dimensionality, and provide the full utility of the data in given context, without the risk of disclosing sensitive raw signal.
Such architecture significantly mitigates a very serious privacy risk related to raw EEG recordings floating around and being used and reused for various purposes.

%% file: sections/introduction.tex
\section*{Introduction}

Electroencephalography (EEG) is a method of recording brain activity as electrical signals, using electrodes placed around the scalp.
The technique has been used for almost a century, with the first historic recording of human brain activity performed in 1924 by Hans Berger~\cite{jung1979fiftieth}.
Since then, the use of EEG has flourished for both research and medical purposes.

Apart from a few notable application areas, such as sleep monitoring~\cite{carskadon2000monitoring}, it is only recently that  EEG has moved outside of the laboratory, with the arrival of low-cost user-oriented neuroheadsets, powerful mobile devices, software frameworks, online services, and methods for data analysis.
Health informatics providers such as Cure4You Technologies\footnote{\url{http://us.cure4you.pro}} are already facilitating storage and interaction with data from health apps.

Datasets of brain activity are being created and made available for analysis and services are starting to be built around EEG data.
While sharing of scientific EEG data is well motivated~\cite{poline2012data}, a strong motivation for sharing may also be present for an individual who acquires EEG data as `self-quantification'.
As EEG analysis is complex and users may be motivated to share data to seek help from the `wisdom of the crowd' for interpreting relations between the EEG and health variability. Or they may use EEG data to enrich and qualify consultation with professionals~\cite{metzger2011using}.
A recent poll made by Pew Internet Projects shows that peer-to-peer health care is already extensive in the US\footnote{\url{http://www.pewinternet.org/Reports/2013/Health-online/Summary-of-Findings.aspx}}.
Professional web services for physicians such as Sermo\footnote{\url{http://www.sermo.com/}} are increasingly quantitative and based on a data sharing.

These development raise questions about proper handling of EEG data and the privacy of users.
Thus the contribution of this article is two-fold.
First, we review privacy issues related to EEG data, caused by the inherent properties of the signal as well as the way it is collected and used.
Second, we propose a framework for controlled sharing of data; acquiring EEG from low-cost mobile neuroheadsets, such as the Smartphone Brain Scanner~\cite{stopczynski2011smartphone}, can be combined with open Personal Data System (openPDS), created for privacy-aware handling of personal data~\cite{de2012trusted} backed by technical and legal means.

\subsection*{Sensitive Use}

Why is there a special need for a  privacy solution in relation to EEG?
In contrast to more conventional sharing of text, imagery, and video, EEG is only partly under voluntary control, hence a user sharing EEG data may only in part comprehend what is being shared.
For conventional data the user can build a mental model of shared content rooted in intuition from everyday social interaction.
This means that sharing may not only be voluntary and transparent, it may in fact be used as efficient personal branding~\cite{labrecque2011online}.
Recent reports on inference of more sensitive hidden variables from conventional social media content, e.g.,~inference of personality factors~\cite{kosinski2013private}, show that such sharing is complex.
But these findings do not rule out that users are aware that the content give away personal characteristics.

\begin{figure}[!ht]
	\begin{center}
	\includegraphics[width=0.6\textwidth]{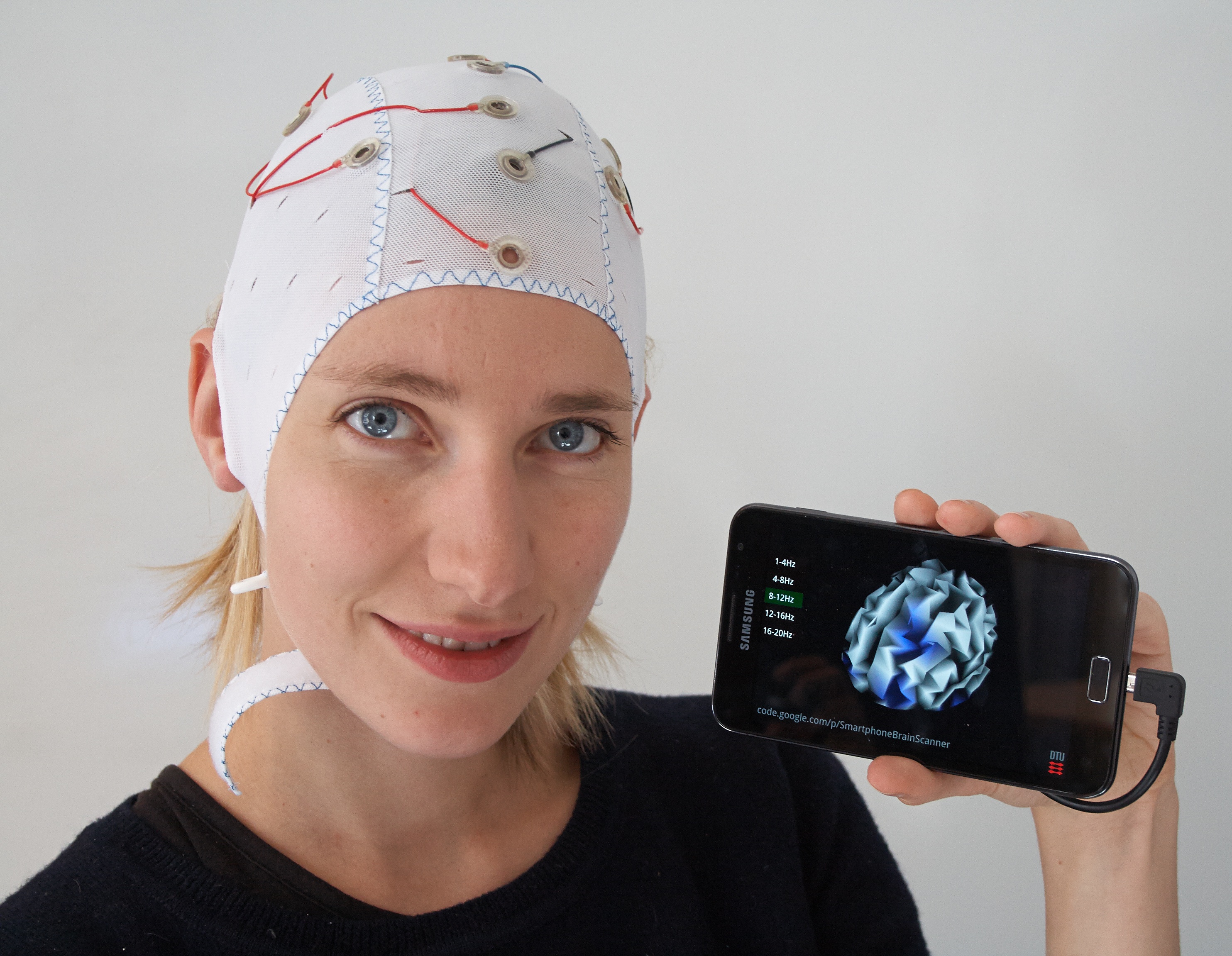}
\end{center}
\caption{
{\bf Mobile EEG brain imaging system, Smartphone Brain Scanner.} Visible in the picture is the entire setup required for data acquisition, processing, and visualization. The cap contains gel-based electrodes used for acquiring electrical singal generated in brain and reaching the scalp. From~\cite{stopczynski2014smartphone}).
}
\label{fig:architecture}
\end{figure}

With EEG data, however, the situation is very different as the major parts of the signal are involuntary.
In fact, the very ability to control minute portions of the variance is the mechanism behind so-called brain-computer interfaces (BCI) where sophisticated machine learning methods are needed to extract the induced components, see e.g.,~\cite{blankertz2007non}.

If we have synchronized, concurrent recordings of EEG and video/behavioral data, the EEG data become grounded, i.e., symbolic in nature, meaning that the information content can be much higher than raw bit rate~\cite{simanova2010identifying}.

What inferences can be made from EEG?
Maybe the most sensitive are related to the diagnostic value of EEG.
EEG has been used for diagnosing various mental diseases.
As early as in 1988, Karson et al.~described the use of EEG for diagnosing schizophrenia, based on increased activity in frequency bands known as delta and beta and decreased activity in the so-called alpha band~\cite{karson1988computerized}.
The study was performed using 20 electrodes on 19 medication-free patients and 21 controls.
Those results have been confirmed and extended in other studies, such as in~\cite{sponheim1994resting}, conducted on 44 first-episode and 58 chronic patients and 102 controls.
The data in this study was collected merely from three electrodes for three minutes.

Dauwels et al.~classified in~\cite{dauwels2010comparative} ``mild cognitive impariment'', a precursor for Alzheimer's disease.
Gotlib et al.~in~\cite{gotlib1998eeg} investigated the asymmetry in the alpha frequency in the frontal region as a possible biomarker for depression.
The recordings were performed from three electrodes for eight minutes (with data cleaned for electrical artifacts as contaminated epochs were deleted).
Many studies have investigated the usage of EEG in depression diagnosis, for an overview see~\cite{davidson2002depression}.

EEG is a well established technique for diagnosing epilepsy~\cite{smith2005eeg}.
Different EEG setups (number of recording electrodes, positioning, recording time, stimulus) are required for different kinds of epilepsy~\cite{lantz2003epileptic, rubin2014efficacy}.
In many cases, however, the basic diagnosis can be obtained with a relatively low number of electrodes (around 20), as typically the seizures affect major part of the cortex.

An important complication of EEG signals is that they are highly personal like fingerprints, DNA, or portraits, thus EEG recordings can be used for identification and authentication of the users.

Revett et al.~\cite{revett2010cognitive} describe cognitive biometrics, utilizing a biosignal approach to user identification.
The EEG signal shows identification accuracy of around 80\%-100\%~\cite{poulos1999neural} when using neural network classifier (LVQ) on the spectral values obtained from the alpha rhythm band of the EEG signal, broken into subbands.
These classification rates were obtained with just two leads (O2 and CZ), and a few  minutes of signal.
Similarly,~\cite{paranjape2001electroencephalogram} reported 100\% classification accuracy for 40 healthy subjects with all data and 80\% using 50-50 split for cross-validation.
These data were obtained from eight electrodes, using around one-minute-long recordings and autoregressive models.
Investigating authentication rather than identification in~\cite{marcel2007person}, Marcel \& Mill{\'a}n reviewed the usefulness of mental tasks for authentication purposes.
This was done using an EEG system with eight electrodes, and machine learning models, on nine subjects.
The performance of the authentication degraded over time (between template recording and authentication attempt), but with data from multiple days overall performance improved.

In~\cite{de2008electroencephalographic} De Gennaro et al.\ showed that humans have an individual profile of the EEG spectra in the 8 - 16Hz frequency during non-rapid eye movement sleep, stable over time and resistant to experimental changes.
The recordings were performed using 19 electrodes.
This indicates the brain activity profile during sleep is highly unique and can be used to fingerprint people.
Their findings were confirmed in~\cite{finelli2001individual}, demonstrating the pattern of the EEG power distribution in non-REM sleep is characteristic for an individual.

Thus, EEG data appear to be highly unique to an individual and thus should be considered extremely sensitive.
The ability to identify subjects in data sets may give the ability to match a short recording of the EEG data with data stored in the large sets, and, if the various types of data are linked, also to link to other information about the user, such as mobility traces or demographics~\cite{de2013unique, sweeney2000simple}.

Using more direct attacks to reveal EEG information, Martinovic et al.~investigated in~\cite{martinovic2012feasibility} how the brain's response to a particular stimulus (so-called P300 paradigm) can be used to narrow down the space of possible values of sensitive information such as PIN numbers, date of birth, or known people.
The tasks required the subject to follow the experimental procedure without explicitly revealing the goal of the experiment: for example thinking about birth date while watching flashing numbers.
Although the presented attacks on the data may not be directly applicable to preexisting EEG data, as they require fairly specific malicious tasks, we can expect --- as the subjects participate in multiple experiments --- correlations violating privacy could be obtained from raw EEG signal.
For example, when a large corpus of the user responses to a visual stimuli is collected, it could be used in P300-based Guilty-Knowledge Test, where the familiar items evoke different responses than similar but unfamiliar items~\cite{abootalebi2009new}.

In~\cite{rosenfeld2006p300}, the authors showed the detection of autobiographical information based on P300 paradigm.
The detection of high-impact, autobiographical information --- possibly more sensitive --- was more reliable than detection of well-rehearsed but low impact, incidental information.
When considering extracting information from the brain activity signal using P300 and related paradigms, the most important pieces may be the ones most easily revealed, invoking the strongest response.

Frank et al.~explored in~\cite{frank2013subliminal} feasibility of subliminal attacks, where the reaction to a short-lasting information of 13.3 milliseconds was measured.
Such stimuli, in theory below conscious perception, could potentially be embedded multiple times in a standard, consciously perceived, stimuli and remain undetected.
Authors showed promising results of recovering whether participants were familiar with a face, analyzing the response evoked by short-lasting stimuli hidden in the video frames.

The brain and the EEG are very far from understood; methods for more accurate analysis of EEG appear on a regular basis.
As we learn to decode more and more advanced cognitive functions, such as the relation between the brain activity and linguistics~\cite{pulvermuller2012meaning}, emotions~\cite{petersen2011smartphones}, or psychological traits~\cite{tran2006personality} it should be clear that we will be able to make unexpected and sensitive inferences from raw EEG signal in the future.
Those who have shared raw EEG publicly are likely to have sensitive personal information lurking in the those data.

We note that in several of the mentioned case studies described above, significant knowledge about an individual has been extracted from relatively short recordings with a low number of electrodes.
Thus, a high number of electrodes and professional grade systems may often not be necessary to `decode' subjects, their mental health, high-level mental processing, and to uniquely identify them.

\subsection*{New Class of EEG Services and Datasets}

Classically, most of the shared EEG datasets have been created as part of scientific experiments with a relatively low number of participants (say, less than a few hundred participants) and without linking the data to other personal data sources.
Such datasets are usually frozen, in a sense that no new data are added to them and during their creation the data were not accessed for the purpose of analysis or building user-facing applications.
Repositories of such datasets can be found, for example, at~\cite{eegdata1} or~\cite{eegdata2}.

This situation is, however, changing, as data start to be collected from larger populations, in a form linked to the individual users, and available for real-time access.
For example, \emph{Emotiv Lifesciences} is a company set up by the creators of the Emotiv EEG neuroheadset with the mission of \emph{``... offering a unique platform for crowd-sourced brain research. Emotiv leverages cloud computing, big data and mobile technology to offer valuable personal insights and accelerate brain research globally.''}~\cite{emotiv}.
MyZeo used to produce an EEG-based headband for sleep monitoring~\cite{shambroom2012validation}, the company does not operate anymore, it however used to allow for data uploading and analysis, providing a service of sleep logging.
Even more companies enter the market, producing the headsets and headbands based on low-density EEG, for example Interaxon producing the Muse band\footnote{\url{http://interaxon.ca/}} or Melon\footnote{\url{http://www.usemelon.com/}}.
Such data do not exist primarily in a form of a frozen, stable dataset, which may be  easier to anonymize, for example using Principle Component Analysis (PCA)~\cite{chaudhuri2013near}.
For the growing data that can be accessed by multiple applications and can be linked with other data sources, the standard anonymization techniques may not be sufficient.

Massive EEG databases containing recordings from thousands of participants are also being build for research purposes.
Brain Resource Database~\cite{gordon2005integrative}\footnote{\url{http://www.brainresource.com/about-us/brain-resource-database}}, integrates information from neuroimaging measures (EEG, ERPs, MRI, and fMRI), arousla (heart rate, respiration, skin conductance responses), neurophysological and persoanlity tests, genomics, demographics.
The database includes the data from over 2,000 normative subjects and number of patients with neurological and psychiatric illnesses.
Another example is Australian EEG Database~\cite{hunter2005australian}, advertised to contain 18,500 EEG recordings and available in a de-identified form.

A great opportunity in linked databases, containing synchronized behavioral and EEG data, is to be able to effectively move from analyzing the weak signals of ongoing free EEG to the much more informative evoked response signals.
In this approach, long-term recordings of EEG data can be augmented with data potentially indicating events that stimulated the brain activations.
In this context, the EEG and other personal data start blending together, allowing for much more complex modeling of human behavior.
Techniques such as parallel factor analysis can be used to extract weak signals from variable responses~\cite{morup2006parallel}.
Achieving a perfect synchronization between EEG signal and behavioral data is very hard; many of the signals collected have naturally different timescales, and the corresponding sensors may not even be able to record with certain resolutions.
In addition, perfect timestamping of the events is difficult, especially on mobile systems that do not provide real-time guarantees.
Shift-invariant multilinear decomposition can be used to analyze such signals,by introducing small adaptive shifts of time series to allow temporal alignment of EEG and behavioral variables~\cite{morup2008shift}.

We are at an early stage of personal data acquisition and sharing, and do not fully understand how the large EEG databases and services will impact the privacy of the participants: How unique users from a general population are in such data, how much can be inferred about the individual, who controls the flow of data and use of the subsequent results.
As EEG analysis methods mature, even more so than what is usually understood as personal data (e.g.~location, friendship graph), access to the raw EEG data will result in very different findings than originally anticipated.
This poses both technical and legal challenges, as the policies developed for datasets, considered to be owned by the researchers or other third parties, such as described in~\cite{eckersley2003neuroscience}, do not fully apply.
Here we argue EEG data should be considered personal data, remaining, as much as possible, under the control of the user.

After all, what is personal if not an individual's thoughts?

\subsection*{New Privacy}

Here we describe how a system for collection of EEG data from low-cost consumer-oriented neuroheadsets, such as the Smartphone Brain Scanner, can be seen as a personal data collection tool and linked to an openPDS backend solution.
In the proposed architecture, the raw data collected by the participants is stored on the server under user control.
The control is enforced by technical (e.g.~self-hosting or encryption) and legal (e.g.~terms of service, contract) means.
The data can be accessed for the purpose of analysis and by user-facing applications, subject to participants' grant of authorization.
Importantly, the raw data are not exposed.
Instead high-level extracted features of the data are only transferred outside of user control as shown in Figure~\ref{fig:architecture}.
This solution promotes the privacy of the user, while at the same time offering the full utility of the data, as additional questions (extracting the high-level features from the raw data) can be installed by the users from the third party services.
It also effectively creates a service offering access to EEG data in a privacy-preserving way.
In many cases, the features extracted from the EEG signal, for example Independent Components (ICs), are of real interest to the researchers or application developers~\cite{onton2006imaging}, and those can be computed in the PDS, under user control.
In fact, recent work suggests that ICs are EEG atoms with a well-defined focal origin that can be used to `explain' their functional roles to the user~\cite{delorme2012independent}.
The user can decide what information is transfered to the third parties, and can better understand what can be done with it.
Multiple PDSes can also communicate with each other in order to calculate an aggregate answer to a question asked to a population, even further increasing the user privacy.

Informed consent of the user to data sharing plays a crucial role in privacy implementation.
As postulated in the Living Informed Consent concept, users should be empowered to understand and make informed decisions about access to their data~\cite{IMM2013-06632}.
The need for better consent procedure has been becoming a widely discussed issue in the biomedical research~\cite{time2012informed}.
For the signal as complex as EEG, claiming that user understands the implications of sharing of the raw data is impossible.
While the extraction of the high-level features, such as spectrograms or ICA components, limits the possible unauthorized uses of the data, it does not significantly change how informed is the user about potential abuses when granting the access.
With access to massive EEG databases we can begin to estimate the effect of features sharing on the possibility of the user identification, providing this calculation before the sharing is executed.
To further improve the understanding, we should aim for the calculation of the highest possible level of the answers in the user-controlled domain; rather than sharing spectrogram of the EEG data, user should share information whether she is epileptic or not.
Only with such level of shared answers, the user can potentially understand the implications of sharing, both the positive and negative ones.

EEG data deserve our attention.
The disclosure of the raw signal can be considered irreversible, as our brain activity remains relatively stable through the life~\cite{dustman1999life} and we cannot replace our brain, at least for now.
As the methods for data analysis and our general understanding of brain increases, recordings obtained once can be re-visited, providing new and unexpected insights.
At the same time, EEG has become very accessible in terms of collection and analysis.
Contrary to other well-established methods of recording brain activity, such as functional magnetic resonance imaging (fMRI) or magnetoencephalography (MEG), EEG can be feasibly used outside of the laboratory and operated by end-users~\cite{Stopczynski2013:SmartphonesAsPocketableLabs}.
Some of the uses of EEG are well understood and attractive for the users, such as accessible brain-computer interfaces (BCI) or neurofeedback applications.
For those reasons, the EEG modality is arguably one of the most sensitive types of personal data that can still be captured in the privacy of home, or even on the go.
Many approaches to sensitive personal data and medical data can be used in the context of EEG; it is important we start a discussion around using such practices in the evolving approaches to EEG data.

It is not a question if the databases of EEG data collected from large populations, for the purpose of providing services and building applications, will be created, but how soon.
For this reason it is essential we begin a discussion around the privacy of EEG data, as seen and treated as personal data, available for public good, aligned with the vision of The New Deal on Data~\cite{pentland2009reality}.

%% file: sections/methods.tex
\section*{Architecture}

Here we outline the architecture of combining the Smartphone Brain Scanner (SBS2) system with open Personal Data System (openPDS).

\begin{figure}[!ht]
	\begin{center}
	\includegraphics[width=1\textwidth]{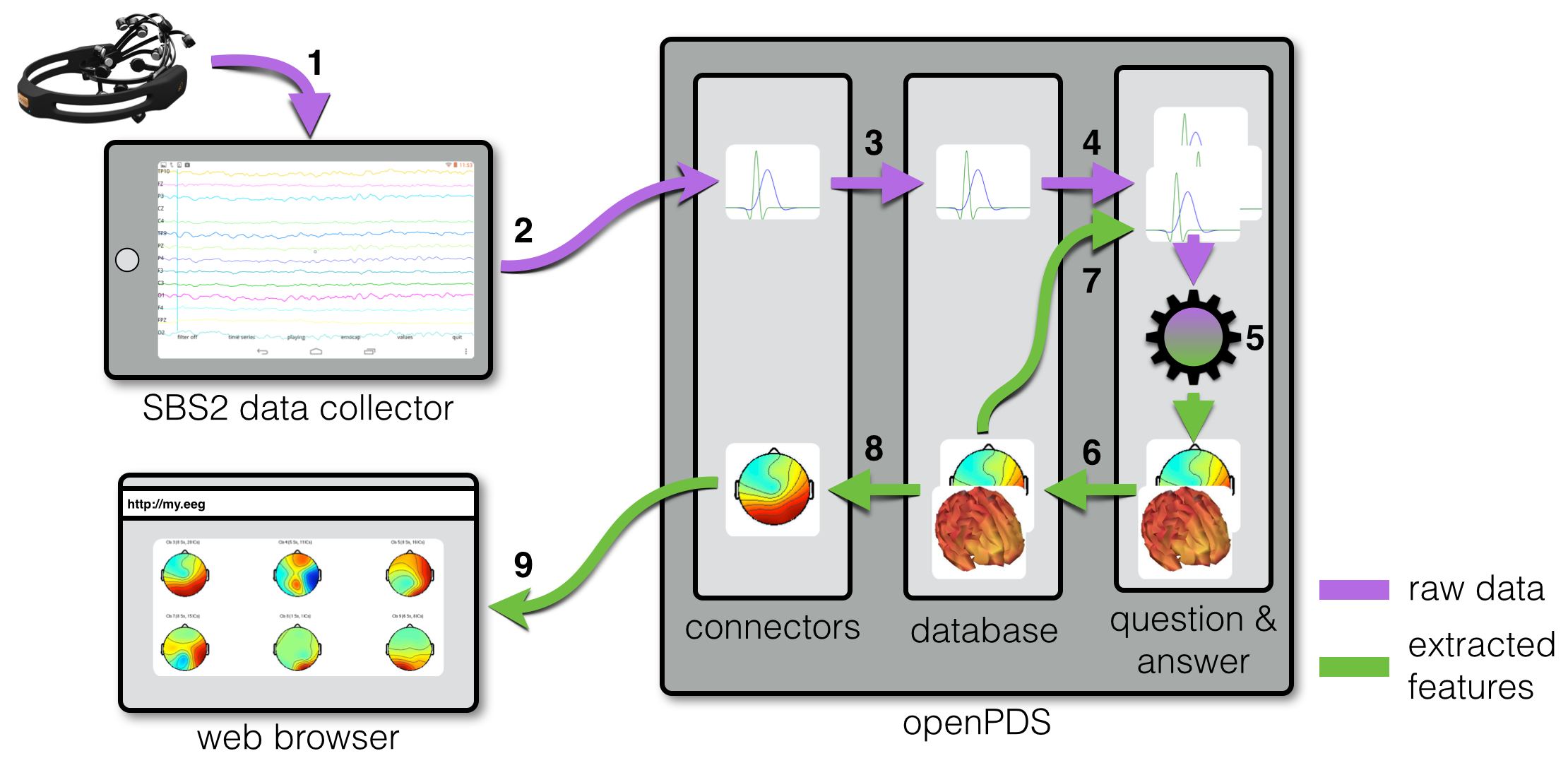}
\end{center}
\caption{
{\bf openPDS integration with Smartphone Brain Scanner.} Raw EEG data are collected from neuroheadset on a mobile or stationary device ({\bf 1}) and uploaded to a server as a binary file ({\bf 2}). Data are then extracted and populated to a database ({\bf 3}). Periodic question \& answer computation process operates on the raw data ({\bf 4}) extracting the high-level features of the data ({\bf 5}). The features are populated into the database in a form of high-level answers ({\bf 6}). Those answers can be used for the computation of other features ({\bf 7}). The pre-computed answers are accessed from the database ({\bf 8}) and served to the requesting application ({\bf 9}).
}
\label{fig:architecture}
\end{figure}

The Smartphone Brain Scanner~\cite{stopczynski2011smartphone} is an open-source system for collecting and processing EEG data from low-cost EEG systems using mobile devices.
The framework has been successfully used to show the reconstruction of the neural sources from emotional stimulus~\cite{petersen2011smartphones}, to implement BCI interaction, and build neurofeedback applications~\cite{Stopczynski2013:SmartphonesAsPocketableLabs}.
The hardware part of the system is based on the off-the-shelf consumer-grade neuroheadsets, such as Emotiv EEG, or custom-built mobile Emocap~\cite{debener2012taking}.
Various platforms and devices can be used for data collection, both mobile (Android) and desktop operating systems (OSX, Linux, Windows).
Primarily used for recordings with mobile devices (smartphones or tablets), SBS2 is an example of advanced personal informatics systems, which can be used not only for research or medical purposes, but also by  end-users.
Full-blown applications can be built on top of the framework, both for data collection and analysis, as well as visualization and feedback.
The recorded data are stored as binary files, containing raw EEG traces and additional metadata (timestamp, user, description, battery levels etc.).

The Personal Data System is a privacy-oriented framework for collection and sharing of personal data~\cite{de2012trusted}.
The particular implementations of the system, developed at MIT Media Lab Human Dynamics group and Technical University of Denmark, are known as openPDS.
The primary feature of openPDS is computation of high-level answers based on raw data and sharing those with other services and applications, rather than exposing the raw data.
Such low-dimensional answers are inherently more privacy-preserving, as they allow the user to better manage and understand what can happen with the shared data.
When raw high-dimensional data are shared, many insights can be gained from them, for example raw GPS traces can be used to infer how much the user exercises, speeds when driving, or nocturnal schedule.
In many cases, sharing such rich data is not required for the service to operate: To get the weather report for the city, users should not have to share their entire mobility trace.

We see the openPDS architecture as a suitable solution for the concerns in sharing personal EEG data.
As described in the Introduction, the EEG data are extremely high-dimensional and can be used to identify users, diagnose mental disorders, or try to extract significant information directly from the recordings.
For those reasons, the sharing of the raw EEG recordings should be as limited as possible.
In the openPDS architecture, the raw data ideally never leave the user-controlled domain, and only the extracted features are shared, based on user authorizations.
Originally created for personal data such as location, transaction records, friendship graphs, etc. the principles of openPDS become even more important in the context of brain activity recordings.

We present the outline of the architecture including SBS2 and openPDS in Figure~\ref{fig:architecture}.
Raw data collected on mobile or stationary device ({\bf 1}) is uploaded to user-controlled openPDS ({\bf 2}) and stored in the database in the raw form ({\bf 3}).
The assumption is that storying the raw data allows for multiple features to be extracted, with the possibility to install more questions in the future.
The uploading application (data collector) has to be authorized by the user (in the OAuth2 sense) to be able to submit the data to her PDS.

A periodic process calculates the answers from the raw data ({\bf 4}), using the algorithms installed by the services that access the data.
The primary reason for periodic calculation of the answers is that those calculations are usually time-consuming and do not necessarily have to be strictly calculated on all the newest data. 
Having the answers readily available when they are requested, is often more important than having the exactly newest answer available.
Nothing however prevents the computations to be performed when the answers are requested, provided such calculations are feasibly fast.

For the computation of the answers, both raw data and previously computed features can be used.
The resulting answers are stored in the database, readily available for sharing with third parties, and to be used internally for other computations within the PDS.
The answers are available as RESTful endpoints, protected by OAuth2 tokens, a solution based on standards common in many Internet-scale services~\cite{de2012trusted}. 
Just as with many other Internet services, users can  authorize third parties to access certain types of data (scopes) from the PDS.
The applications accessing the data can live in the web-browser, on mobile devices, or as standalone programs.

The openPDS is not limited to storying only brain activity recordings of the user.
Considering the EEG recordings as another personal data, that should be under the same user controls and shared in the same way, makes it easy to mash up the data from different sources.
In a simple case, the additional data can be seen as metadata for the EEG recordings.
For example, every time the user captures her brain activity, location can be saved and uploaded accordingly.
An example answer that can be computed from such data is a list of places where user tends to get drowsy, without revealing the raw EEG recordings or the exact mobility traces.
In more complex cases, where multiple types of data are collected, the EEG recordings become yet another data source that can be used for modeling of the user.

\begin{figure}[!ht]
	\begin{center}
	\includegraphics[width=0.8\textwidth]{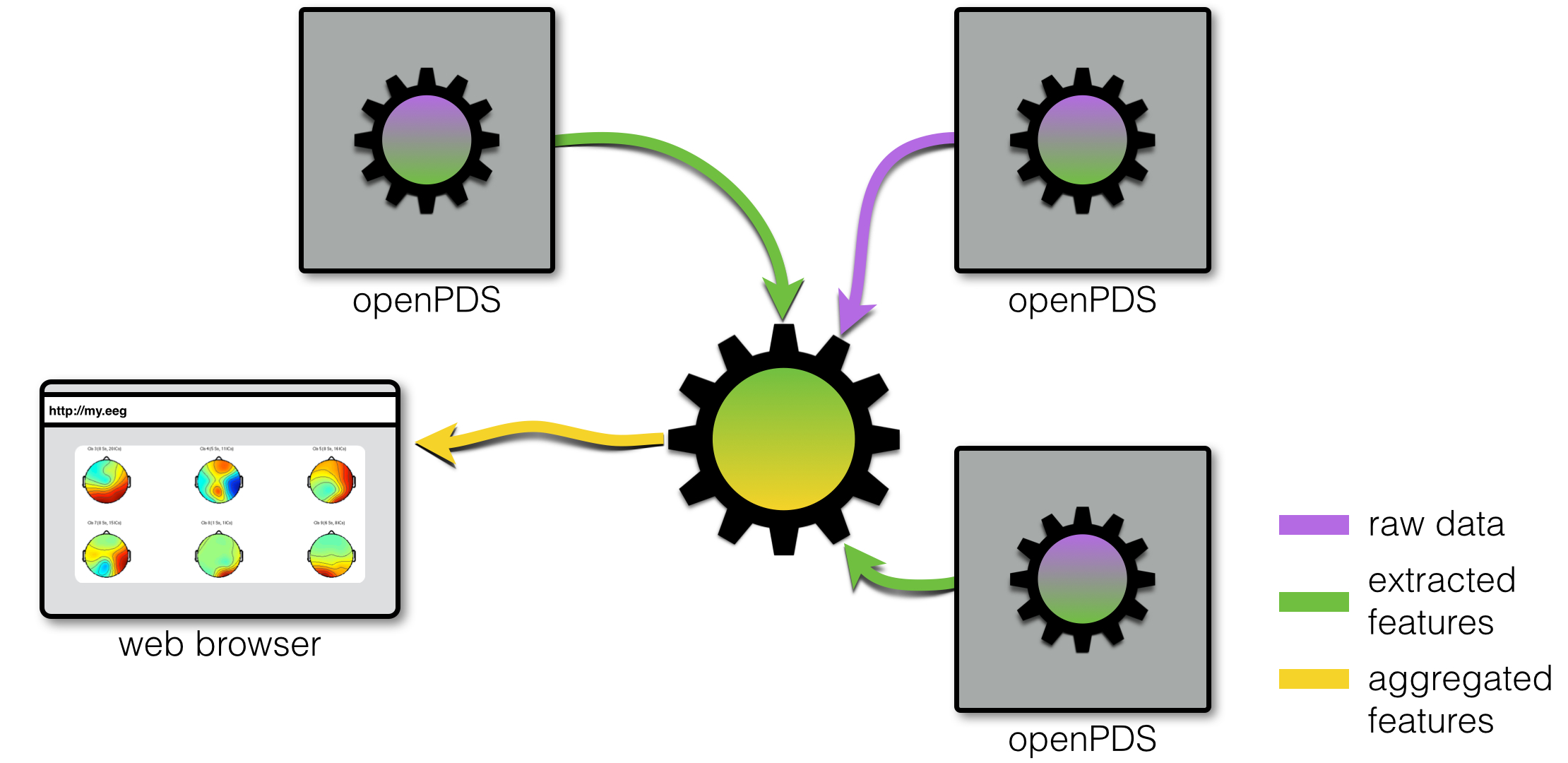}
\end{center}
\caption{
{\bf Group computation of answers from multiple PDSes.} Depending on the nature of the computation, individual users' PDSes can contribute raw data or extracted features, and aggregate answer is available to the calling service.
}
\label{fig:architecture_group}
\end{figure}

The solution presented here promotes the end-user control over sensitive data, at the same time making these data readily available for research purposes.
Importantly, the architecture allows for privacy-preserving access to the data in real-time, making it possible to build services and applications on top of it.
Additionally, the computations can be aggregated, in that the answers are computed from a group of PDSes, as presented in Figure~\ref{fig:architecture_group}.
This can provide insights about the state of the population rather than individuals, potentially even more valuable for research purposes and privacy at the same time.
Certain of those aggregation computations between PDSes can be performed in a privacy-preserving way, where no single entity learns the entire dataset.
For example, collaborative filtering used in recommendation applications, can be done in a privacy-preserving way, where no information is leaked between the nodes participating in the computation~\cite{canny2002collaborative}.
Similarly, support vector machine (SVM) classification, one of the most popular classification methods for data mining and machine learning, can be performed under certain assumptions without disclosing the data of each party to others~\cite{yu2006privacy}.
Comprehensive overviews of privacy-preserving machine learning methods is presented in~\cite{agrawal2000privacy} and~\cite{aggarwal2008general}.
It is outside of the scope of this article to investigate the particular solutions of the privacy-preserving machine learning, as those heavily depend on the application; here we signal the existence of the solutions potentially applicable for the widely-used treatment of EEG data.

OpenPDS for EEG data should offer several core features to effectively improve privacy controls.
Primarily, openPDS needs to offer storage of structured data, accessible via API.
The structure in the data is important for enabling the major feature of openPDS, computation of answers.
OpenPDS is not a storage of unstructured data, like for example Dropbox, but offers execution of the calculations.
Sharing of the calculated answers should be strongly preferred over sharing of the raw data; it may even be enforced that raw data is never shared and only sufficiently privacy-preserving features leave user-controlled system.
The sharing of the data should be realized as a scalable service, based on user authorizations and using standards, such as OAuth2.
All the flows of data should be realized through this API, without backdoor access: healthcare providers, health applications, or researchers all should use the same mechanism.
The data access must be audited, so the user can monitor the data flows and either by manually accessing the logs or---preferably---by using tools automating the process by notifying only about unusual or suspicious data accesses.
PDS implementation must be supported by a well-designed user interface, making the sharing and monitoring actions comprehensible for the end-user.
Finally, the technical solutions of openPDS must be backed with aligned legal and business terms, supporting roles of the entities, their obligations, and allowed data flows.

%% file: sections/legal.tex
\section*{Legal Framework}

Brain activity datasets of the types described in this article pose especially important legal and policy issues.
Personal data carry with them a wide variety of obligations and rights in general.
The potential of neurological data to uniquely identify a person and to detect and convey psychological or other medical conditions raises particularly sensitive legal and public policy issues.
The legal framework applicable to neurological data will determine which rules apply to these issues, and therefore what legal outcomes will occur.
The relevant facts and circumstances surrounding the neurological data are the basis for establishing which legal frameworks are applicable.  

\subsection*{Roles and Relationships}

Whether a party is a data controller, a data holder, a data custodian, a data agent, a data receiver, or a data processor---to make but a few legal roles---will depend largely upon the underlying facts and circumstances of their particular involvement with brain activity data.  Some parties will play combinations of various roles, and some parties will engage in just a few of the functions allocated to a given role, perhaps as an outsourced service provider the the party that is chiefly responsible in the role.
From a technical perspective, each role carries with it a span of functions and expected interactions with other roles.  From a legal perspective, each role has corresponding rights, obligation, and other applicable rules for the role.  

The law can only be understood when it is factored into identified situations and contexts.
One key facet of relevant context arises from the roles and relationships of any other individual or organization involved with the neurological data.
Consider, for instance, a situation involving the measurement of a patient's brain activity by their medical doctor.
Clinical data used in a clinical context will likely invoke legal frameworks with detailed and prescriptive requirements about conduct like data collection, information security, and soliciting consent, such as Health Insurance Portability and Accountability Act (HIPAA) or Doctor/Patient Confidentiality.
In such situations, other potentially applicable frameworks can include rules governing standards and quality of medical care (Malpractice) or limits on cost of service and billing practices (CMS Medicare/Medicaid) or even the contractual terms and intellectual property rights to information created as part of a medical encounter (patent).  

It is difficult to conceive high stakes legal issues arising if no other person or organization has any role with or relationship to a given individual's creation, use, and deletion of their own neurological data.
In theory, a purely and exclusively individual scenario of use can be imagined, assuming the brain activity equipment and the resulting neurological datasets are of, by, and for the same individual and no other party has any relationship, rights, responsibilities, role, or point of interaction interaction whatsoever.
In this case, neither the raw data nor derived data would be shared or otherwise accessed by any other person and no basis for privacy issues would seem plausible.  

By contrast, many potential legal frameworks may be triggered if that same individual provides the same neurological dataset to another legal entity.
The legal obligations on the receiving party may vary based upon whether the individual was compelled to reveal, formally consented to provide, or informally choose to share their brain activity data.
If duress or coercion compelled the data subject can invalidate consent and even undo contractual agreements.
Furthermore, the rules may vary significantly depending on whether the individual disclosing their neurological data was under 18 years of age, e.g.~\cite{ageoflegal} , actively serving in the military~\cite{ucmj,geneva}, was sleep walking at the time, had been lied to about the nature of the data\footnote{\url{http://www.ftc.gov/ftc-policy-statement-on-deception}} or many, many other factors bearing on the capacity of the individual to make sound decisions about disclosure.  

The roles may depend and change depending on the residence of the participant.
For example, if the neurological data is of the brain activity of a resident of Massachusetts, additional roles may apply with a corresponding layer of relationships and information security obligations.
The Massachusetts General Laws establish a statutory and regulatory scheme requiring service providers to encrypt personal information about a resident of the state, among other requirements (M.G.L. c. 93H, and 201 CMR 17.00).
These rules apply when the brain activity data is associated with certain other personal information used to identify the user or as part of a billing relationship for a service (201 CMR 1702 Definitions: Personal information).   

Under this Massachusetts legal framework for personal data information security, the key roles are (201 CMR 1702 Definitions: Service provider and Person): 
\begin{itemize}
\item Service provider---any person that receives, stores, maintains, processes, or otherwise is permitted access to personal information through its provision of services directly to a person that is subject to this regulation
\item Person---a natural person, corporation, association, partnership, or other legal entity, other than  an agency, executive office, department, board, commission, bureau, division or authority of the Commonwealth, or any of its branches, or any political subdivision thereof
\end{itemize}

If the same data were held by a department or other unit of the state government, then the Massachusetts Fair Information Practices Act may apply yet a different layer of relationships and a range of respective duties and particular work flow for data access, record keeping and consent based sharing (M.G.L. c. 66A, informally known as FIPA\footnote{\url{https://malegislature.gov/Laws/GeneralLaws/PartI/TitleX/Chapter66A}}).
The key rights and responsibilities are very similar to those proclaimed by the New Deal on Data~\cite{pentland2009reality} (see below), including a legislated right for people to be informed of the personal data about them held by the state, to be told of any third party access to that data and the purpose for that access, to request and receive a copy of the personal data about them, and to ensure that data is not shared with other parties unless they personally consent to each such disclosure.  

The key roles under the Massachusetts FIPA legal framework are: 

\begin{itemize}
\item Agency---any agency of the executive branch of the government, including but not limited to any constitutional or other office, executive office, department, division, bureau, board, commission or committee thereof; or any authority created by the general court to serve a public purpose, having either statewide or local jurisdiction
\item Data subject---an individual to whom personal data refers, not including corporations, corporate trusts, partnerships, limited partnerships, trusts, nor other similar entities
\item Holder---an agency which collects, uses, maintains or disseminates personal data or any person or entity which contracts or has an arrangement with an agency whereby it holds personal data as part or as a result of performing a governmental or public function or purpose. A holder which is not an agency is a holder, and subject to the provisions of this chapter, only with respect to personal data so held under contract or arrangement with an agency
\end{itemize}
 
As will be discussed later in this section, the roles and duties of parties depend upon how personal data is defined in the context applicable to those parties.  For instance, the definition of personal data under the Massachusetts FIPA law is: 

\begin{quote}
{\it
any information concerning an individual which, because of name, identifying number, mark or description can be readily associated with a particular individual; provided, however, that such information is not contained in a public record, as defined in clause Twenty-sixth of section seven of chapter four and shall not include intelligence information, evaluative information or criminal offender record information as defined in section one hundred and sixty-seven of chapter six. 
}
\end{quote}

Whether information is contained in a public record or not does matter under this definition.
The United States Supreme Court has indicated that public records relating to a person, when taken together from many different sources, may constitute in the aggregate a violation of privacy rights.
Under this standard, despite the public record status of data, it is possible that it will nonetheless be deemed to be personal data under the law.
Therefore, whether a given party is or is not in the role of a personal data holder does depend upon how personal data is currently defined in a given context.
Parties that generate, receive, store, analyze, and share brain activity data are likely to exist in several contexts and therefore to hold a variety of roles with respect to that personal data.  

Clarity is needed regarding the role of the parties initially engaged in the provision and use or hosting of brain activity obtained with consumer level equipment and data storage enabling individuals to generate and use such data about themselves.
The relationship between the individual data subject and the company or companies providing the equipment and services needed to create and use brain activity data in an individual or small group consumer context will determine the legal results and privacy rights, responsibilities, and other rules.
The role and relationships between individuals and providers of personal neurological equipment and services can be found in the contracts and other agreements between those parties.
The terms of service, privacy policy, and other such agreements literally and explicitly define and describe the roles and obligations of each party vis-a-vis each other party.
Industry practices, standard,s and common approaches are needed to ensure widely understood terms and conditions are consistently applied.
While statutes and regulations can provide further certainty about the legal roles and relationships of parties to brain activity data, the use of common and agreed contractual terms is a more agile and adaptable method.  

\subsection*{Reasonable Expectation}

Parties who play a role in the use of brain activity data must respect privacy interests of of the data subject.
But defining the appropriate individual rights and obligations allocated to each role depends on the privacy and related frameworks applicable to those roles.
Fundamentally, the law reflects agreed or at least widely understood expectations about behaviors and consequences.
Some statutes explicitly base a rule on whether a behavior would be considered an ``unreasonable'' interference with a right under all the relevant circumstances.
For instance, state law of the Commonwealth of Massachusetts provides: ``A person shall have a right against unreasonable, substantial or serious interference with his privacy'' (MGL Ch214 Sec1b).
However, precisely what behavior or situations are deemed reasonable or unreasonable are deliberately left to adjudication on a case by case basis.
Naturally, as cultural, social, political, and other norms change, the line between permitted legal conduct and prohibited privacy violations will change correspondingly.  

The trends toward open data, quantified self, and social networking are gaining momentum. 
These changing attitudes and practices are also moving the set-point for what types of conduct might be reasonably agreed to be privacy violations.

\subsection*{The New Deal on Data}

There are many existing legal frameworks covering personal data.
The novelty and quickly evolving nature of products, services, and use cases centered upon personally created and used neurological readings presents unprecedented factual contexts and therefore yields uncertainty about the applicable rules.
For this reason, and also in order to ensure the value of these creative technologies remains available and grows, it is important to establish a sound and predictable legal framework applicable to personal neurological data and surrounding practices.  

The New Deal on Data~\cite{pentland2009reality} provides a simple, efficient and effective approach for establishing the legal framework for personal neuroinformatics.
The New Deal on Data is a refined and focused statement of the most fundamental facets of the fair information practices:

\begin{enumerate}
\item You have a right to possess your data. Companies should adopt the role of a bank account for your data, where you open an account (anonymously, if possible), and you can remove your data whenever you like.

\item You, the data owner, must have full control over the use of your data. If you are not happy with the way a company uses your data, you can remove it. All of it. Everything must be opt-in, and not only clearly explained in plain language, but with regular reminders about the status.

\item You have a right to dispose or distribute your data. If you want to destroy it or remove it and redeploy it elsewhere, you have the right to do it.
\end{enumerate}

\subsection*{The Overarching Evolving International Legal Framework}

The 2013 Organization for Economic Co-operation and Development (OECD) Privacy Guidelines\footnote{\url{http://www.oecd.org/sti/ieconomy/2013-oecd-privacy-guidelines.pdf}} represent the first revision to the OECD fair information practices since they were initially agreed internationally in 1980.

An important aspect of these Guidelines is their focus on obligations of those who are Data Controllers meaning ``a party who, according to national law, is competent to decide about the contents and use of personal data regardless of whether or not such data are collected, stored, processed or disseminated by that party or by an agent on its behalf''. 
The Data Controller is obliged to protect Personal Data under their control, and such data is defined as ``any information relating to an identified or identifiable individual (data subject)''.
This includes an obligation to keep personal data secure and correct and to provide, upon the request, an individual with a copy of the personal data about them.
These guidelines are intended to apply to both public sector and private sector Data Controllers.  

A major update to the OECD Privacy Guidelines includes some clear signals about top priority legal and policy reforms that are highly relevant to neurological personal data.
Two of the most important topics identified in the updated text are:

\begin{enumerate}
\item Biologically based human information, including biometric and genomic data and how this bio-info is an important emerging class of personal data with unique legal implications.
\item Big Data and statistical models, including predictive analytics as a harbinger of a very different playing field for privacy and broadly held expectations about the nature and purposes of personal data flows.
\end{enumerate}

Regarding ``the human body as information'' the 2013 OECD Guidelines note: 

\begin{quote}
{\it
Advances in genetic technology have important implications for the health of individuals, helping researchers better understand, prevent and treat various diseases. Genetic testing to assess health risks or to determine biological relationships raises issues that affect not only an individual’s privacy but also raise the issue of `group privacy', as our genetic makeup is shared by other members of our family and community. At the same time the indelible nature of genetic information and its potential implications for discriminatory treatment make it particularly sensitive.
}

{\it
Commonly viewed as a means of identification and authentication, biometrical information is beginning to be collected and used in a greater variety of contexts–--from voice recognition systems for allowing employees to access business applications to digital fingerprinting to pay for lunch at an elementary school. 
As technology advances, the use of additional human characteristics as information will continue to pose challenges to notions of privacy and dignity. The reliability of biometric information and systems has improved, and biometrics are generally considered strong and valuable to authentication systems. The question of whether biometrics invades privacy or protects it, or both, as well as the appropriateness of relying on biometrics to resolve problems or make decisions about individuals, will be issues that will need to be considered as biometric technologies evolve.
}

\end{quote}

The direction of a policy and legal direction in alignment with the principles of the New Deal on Data is even clearer when the commentary regarding human biological data is read in the context of global trends toward adoption of and reliance on Big Data and data-driven services and lines of business.
In relevant part, the 2013 OECD Guidelines observe: 

\begin{quote}
{\it
The development and use of algorithms and analytics has made large data sets more accessible and capable of being linked, which can result in increased and new uses of the data, thereby making data more valuable. The remarkable pace of development and evolution of technologies and business models make it less easy to accurately describe potential future uses of information at the time of collection. This has resulted in a desire to keep personal data for an as-yet undefined, later purpose and reflects the intrinsic value of personal data to both business and governments. Search engines, which allow for easy, global searches of any personal data made public, make data retrieval much easier for Internet users. Growing use of linked data sources and contextual semantic technologies allow for greater and more sophisticated automation in the discovery and aggregation of personal data. Automated decision-making through data mining and rule engines is increasingly possible in a variety of contexts. Moreover, searches are no longer restricted to text and numbers: facial recognition applications now allow users to identify individuals in images online with growing accuracy. The phenomenon of ``big data'', namely, the vast quantities of data that can be stored, linked, and analysed, brings with it the possibility of finding information, trends, insights that were not previously obvious or capable of being ascertained. This may hold great economic and social value, but there can be privacy implications.
}
\end{quote}

Understanding the role of every party to the creation, use, access, modification, sharing, and destruction of brain activity data is key to applying an acceptable legal framework.
If the commercial company providing consumer equipment and services needed to collect personal data brain activity is considered a Data Controller in the OECD Privacy Guidelines sense, then a New Deal on Data will follow for down stream uses and contexts.
However, if such providers are considered eCommerce-like owners of services and data systems in the model of today, then very different legal outcomes will likely follow.
Whether individuals are an immediate and continuing role as owners or at least key control points over their brain activity is essentially a question of which legal framework roles and relationships will be applied.  

In the aftermath of major defining security failures, from the Snowden disclosures to the Target breach to name only two, there appears to be a rare opportunity for deeper and broader legal and policy reform than has been witnessed in many years.
Many US state legislatures are debating statutes that would prevent and/or severely punish personal data abuses while the EU is increasing pressure to repeal the long-standing ``Safe Harbor'' agreements for trans-Atlantic personal data flows in response to the evident lack of personal data stewardship on the Western edge of the partnership.
The National Strategy for Trusted Identity in Cyberspace (NSTIC) Identity Ecosystem Steering Group (IDESG) is one example of a potential avenue for fresh thinking from a shared set of basic values founded on the Fair Information Practice Principles.
This type of multi-stakeholder forum on personal data standard and policy framework could validly develop, credibly propose, and provide continuing support for New Deal on Data oriented identity data frameworks.  

%% file: sections/discussion.tex
\section*{Discussion}

In the biomedical field, there is a growing discussion about how informed consent and data sharing practices are in need of serious improvement~\cite{erika2012informed, time2012informed}.
It would be irresponsible to continue collection of data of higher and higher resolution, from growing number of participants, over long periods of time without the discussion about how to provide better privacy guarantees.
This is especially true for the biomedical data, that change little in persons' lifetime and once acquired by malicious parties can do significant harm for a long time.

The main goal of building privacy-preserving services for personal data is not to hide the data or to make them unavailable; quite the opposite.
We need more data sharing for the public good, EEG recordings are no exception.
Implementation of end-user control over the data is a way of increasing data liquidity, allowing for more organized and better managed flows.
With cheap recording devices and online services able to generate value for the end-user, for the first time in the history we can start looking at the brain activity of the entire populations.
It is however important that such data will not become exclusive to commercial services, closed in silos unavailable for large-scale research.
Implementation of the architectures such as one outlined here is a way to promote data availability while protecting the users contributing these data.
This is well aligned wit the concept of the New Deal on Data~\cite{pentland2009reality}, postulating increased availability of the personal data driven by end-user data ownership.

We postulate the data ownership should be given to the user, at the same time recognizing that the EEG data is extremely complex; even short recordings can be useful for many purposes.
The only sensible way to increase the data availability while protecting the privacy of the users is with the question \& answer mechanism.
Very significant portions of the calculations must happen under user control, when only extracted features are shared with the third parties; features that make it possible to understand what knowledge can be extracted from them.
The technical solution of question \& answer will not be perfect. 
Even when sharing very high-level features, there are still dangers of abuse: multiple answers can be combined, sensitive answers can be shared without user authorization, new analysis methods can allow for reuse of the shared features.
Researching how to limit those on the technical grounds, for example by monitoring how the requested answers cover the original signal, is important but not sufficient.
The legal framework, including contract governance, credible threat of legal consequences, and robust auditing system need to be integrated in the systems.
At the end of the day, if there is money to be made from the data abuse, technical means will be defeated by motivated attacker and only legal framework can limit a widespread abuse.

Significantly more research about the sensitivity of the EEG data is urgently needed.
If I post a minute of my raw EEG data on Facebook today, will I become indefinitely identifiable in every subsequent EEG database?
Will the researchers of tomorrow be able to learn about my mental diseases?
Without even rough answer to those questions, it is very hard to discuss and implement best practices for handling personal EEG data. 

\begin{figure}[!ht]
	\begin{center}
	\includegraphics[width=0.7\textwidth]{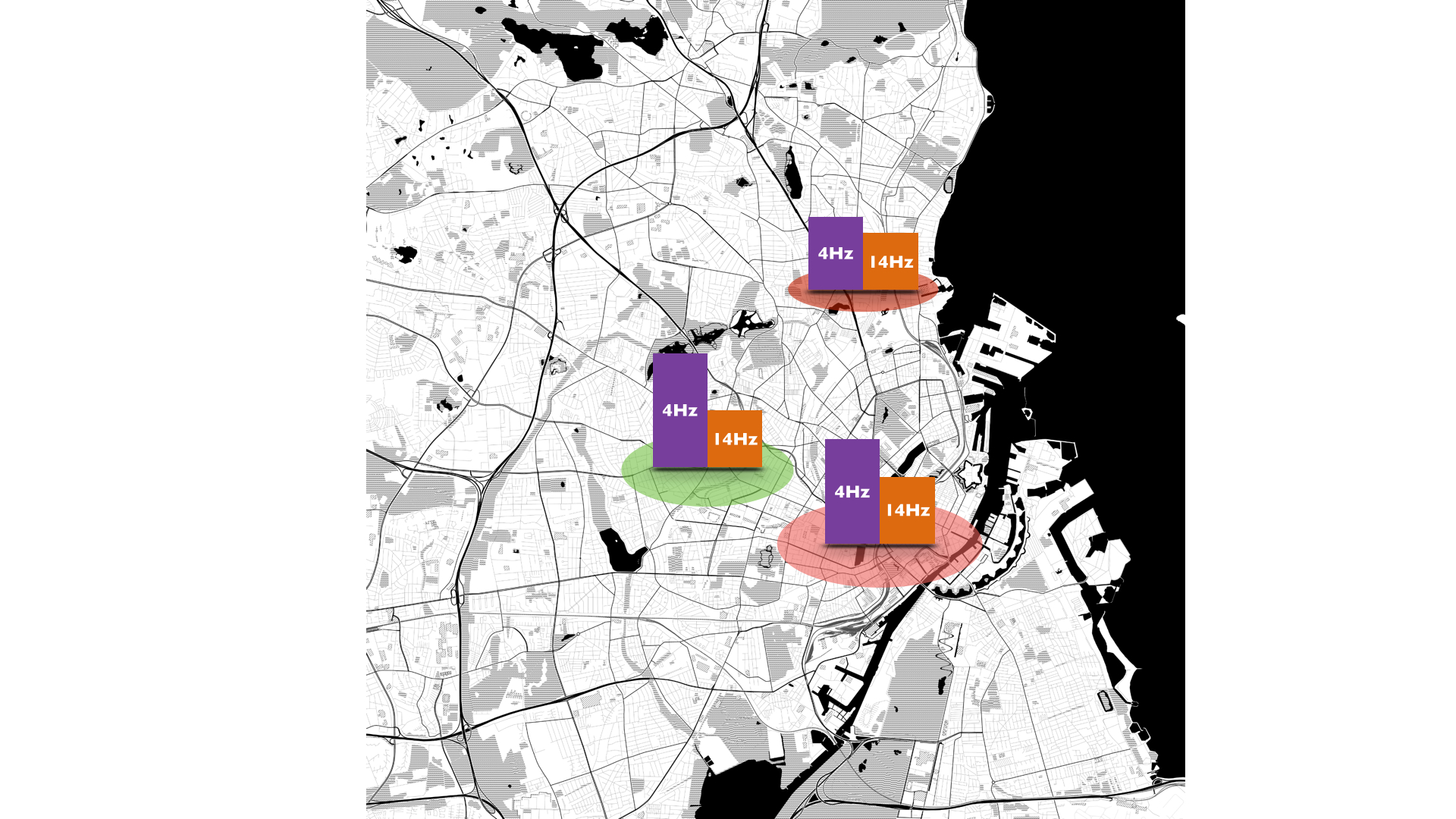}
\end{center}
\caption{
{\bf Mockup of a service showing geo-tagged brain activity frequencies.} Frequencies displayed are recorded by users using personal neuroinformatics system, such as Smartphone Brain Scanner, and associated with location data. Researchers may access view aggregated over multiple users and time, whereas users can see their own data exactly.
}
\label{fig:map_mockup}
\end{figure}

As we build online services for collection and analysis for EEG data and deploy research studies with unprecedented capabilities of EEG recording, very novel value will be available for service providers, researchers, and users.
For example, in Figure~\ref{fig:map_mockup} we show a mockup of a service showing geo-tagged results from brain scans.
Showed frequencies, 4 and 14 Hz have been associated with drowsiness level~\cite{jung1997estimating}, and such map could be a service for plotting the engaging places in the city.
Or, if applied to scans performed while driving a car, a live monitoring tool for mapping places and times, where the drivers become dangerously drowsy.
Such services can be possible with the development of 24/7 EEG recording methods, for example low-dimensionality neuroheadsets, subcutaneously placed electrodes~\cite{beck2007method}, or electrodes placed in the ear canal~\cite{looney2012ear}.
Researchers and service providers in openPDS architecture may only access aggregate data from multiple users, averaged over time, and only certain features.

Services collecting and processing massive biomedical and health data---including EEG---should adapt the openPDS approach, offering to the user hosting of their data, with the understanding that users can control the data access authorizations, request deletion of the data, or move the data to another service provider.
Clear boundaries within those services should be set, defining in business, legal, and technical aspects what is under user control and what extracted high-level answers are used for providing the services.
Business model of collecting large datasets from the users in exchange for a service, and subsequent selling access to those datasets to the researchers is arguably a dangerous model in a context where the sensitivity, value, and proper anonymization techniques are not sufficiently researched.
It would be a broken economy.

Here we presented an outline of a solution, one way of providing privacy for personal neuroinformatics.
Many questions still need be asked and answered.
What are the precise legal frameworks for treating high-resolution biomedical data as personal data.
What are the features and answers that can be considered safe to share.
Are ICA components such answers?
Source reconstructions of the activity?
Spectrograms?
What can be used to identify the users and how well, or what unexpected findings can be computed?

We hope to invite the neuroscience and EEG communities to discuss the privacy and liquidity of the data, as seen in the context of online service and massive research studies.
The time of personal neuroinformatics is coming, and such discussion is necessary before we end up with extremely sensitive data floating around wildly.
Fixing this a posteriori may be difficult, if not impossible.
We should own our brain activity, an extremely valuable and sensitive asset that we should have the right to contribute for the public good.